\documentclass[aps,prb,floatfix,amsmath,amssymb,superscriptaddress,twocolumn,nofootinbib]{revtex4-2}
\usepackage{bm}                     
\usepackage{appendix}
\usepackage[latin1]{inputenc}
\usepackage{dcolumn}      
\usepackage{bbm}
\usepackage{color}
\usepackage{xcolor}
\usepackage{esint}

\usepackage[mathscr]{eucal}
\usepackage{latexsym}
\usepackage{amsbsy}
\usepackage{float}
\usepackage{graphicx}
\usepackage{epsfig}
\usepackage{subfigure}
\usepackage{wrapfig}
\usepackage{epsf}
\usepackage{float}
\usepackage[normalem]{ulem}
\usepackage{qcircuit}
\usepackage{diagbox}

\definecolor{darkblue}{HTML}{004D6B}
\definecolor{darkred}{HTML}{8c1515}
\definecolor{darkgreen}{HTML}{006400}
\usepackage{hyperref}
\hypersetup{ colorlinks=true, urlcolor=darkblue, citecolor=darkred,
    linkcolor=darkblue, breaklinks }
\usepackage{parskip}

\newcommand{\be}{\begin{equation}}
\newcommand{\ee}{\end{equation}}
\newcommand{\ba}{\begin{array}{l}}
\newcommand{\ea}{\end{array}}
\newcommand{\re}[1]{(\ref{#1})}
\newcommand{\ci}[1]{\cite{#1}}
\newcommand{\banonum}{\begin{eqnarray*}}
\newcommand{\eanonum}{\end{eqnarray*}}
\newcommand{\baa}{\begin{eqnarray}}
\newcommand{\eaa}{\end{eqnarray}}
\newcommand{\bfr}{\begin{flushright}}
\newcommand{\efr}{\end{flushright}}
\newcommand{\bfl}{\begin{flushleft}}
\newcommand{\efl}{\end{flushleft}}
\newcommand{\lab}[1]{\label{#1}}

\def\DM#1{{\color{red}Davron: #1}}

\begin{document}
\title{One-dimensional relativistic hydrogen-like atom in Dirac materials:\\
Energy spectra and supercritical states}

\author{S.Z. Rakhmanov}
\affiliation{Chirchik State Pedagogical University, 104 Amur Temur Str., 111700 Chirchik, Uzbekistan}
\affiliation{Cyber University, 42 Yangiabad Str, 111500 Nurafshan, Uzbekistan}
\author{K.P. Matchonov}
\affiliation{National University of Uzbekistan, 4 University Str., 100174 Tashkent, Uzbekistan}
\author{A.K. Rakhimov}
\affiliation{Chirchik State Pedagogical University, 104 Amur Temur Str., 111700 Chirchik, Uzbekistan}
\author{D.U. Matrasulov}
\affiliation{Turin Polytechnic University in Tashkent, 17 Niyazov Str., 100095 Tashkent, Uzbekistan}

\begin{abstract} 
We consider a model of 1D relativistic hydrogen-like atom, formed by a Coulomb impurity in graphene nanoribbon. Describing the electron motion in  terms of the one-dimensional  Dirac equation for Coulomb potential taking into account the finite-size of the atomic nucleus, we compute the eigenvalues and eigenfunctions of the atomic electron. 
The cases of unconfined atom and atom-in-box system are considered.  Special focus is given calculation of supercritical energy levels and the critical charge. The latter is the value of the atomic nucleus charge, when the electronic state reaches the border of the Dirac sea.  It is found that for confined atom the value of the critical charge is larger than that of free atom. Experimentally measurable characteristics, local density of states is also plotted for both cases. Existence of strong localization for atom-in-box system is shown.
\end{abstract}

\maketitle

\section{Introduction}
Charged impurities in graphene can form low-dimensional atoms where electron motion are described in terms the Dirac equation, i.e. electrons in such atom  can "mimic" relativistic behavior in one- or two dimensions. Due to this fact atoms formed by Coulomb impurities provide effective tool for the study of relativistic phenoemna, including QED in low-dimensions at the "table- top" level.  This feature makes possible to study relativistic physics and high-energy phenomena   without using big machines (accelerators of heavy ions) i.e. using low-cost facilities. Relativistic quantum mechanichs and QED with high-Z ions in $(3+1)-$ dimensions have been subject for extensive research during past  five decades, both in theoretical \cite{HiZ0,Greiner85,Popov01} and experimental contexts \cite{Backe78,Kozhuharov79}. The latter requires using high-cost facilities, including accelerators of high $Z$ ions \cite{HiZ1,HiZ2,HiZ3,HiZ4}. 
Remarkable phenomena which can be studied in the context of QED in $(3+1)-$ dimensions and relativistic quantum mechanics are so-called critical and supercritical phenomena, such as electron-positron pair creation, vacuum polarization and Klein paradox \cite{Greiner85,Popov01,Bawin81,Popov20}. However, experimental realization of such processes are rather high-cost "game", as they require using rather big accelerators. Mapping them onto the graphene and studying in two dimensions makes such experiments very cheap (by 3 -4 orders). In addition, even relativistic quantum mechanics in two dimensions in vacuum is much simple that greatly facilitates some aspects of studies of these phenomena (e.g., the critical Coulomb charge in 2D case two times smaller than that in $3D$ \cite{Khalilov98,Khalilov17}). In low-energy regime Coulomb impurities in graphene can be described in terms of two-dimensional Dirac equation. Coulomb problem in two dimensions was studied first in the Refs.~\cite{Khalilov98,Khalilov17}), where the problem of critical charge was also considered. Electronic structure and  scattering in a model of 2D relativistic atom with screened Coulomb potential in graphene was considered in detail in the Ref.~\cite{Novikov07}.
The (planar) Dirac equation based theoretical studies of  the  Coulomb impurities in grahene was considered first in the Refs.~\cite{Biswas07,Shytov,Pereira07,Pereira08,Gamayun09,Martino14,Kloepfer14,Babajanov14,Luican14,Wang12,Wang13,Rakhmanov24_01,Rakhmanov24_02}. Experimental studies of supercritical phenomena in such systems were p[resented in the  \cite{Wang13,Wang12,Luican14}. 
Coulomb problem for graphene with supercritical impurity was considered in analogy with the $3D$ counterpart in   \cite{Kuleshov17,Popov20}. Field induced ultrafast phenomena, including electron-hole pair creation in supercritical impurities in graphene have been studied in the Refs.~\cite{Rakhmanov24_01,Rakhmanov24_02}.  Despite such a progress, all the studies are still restricted by two-dimensional atoms, although Coulomb impurities in graphene allow to create and study one-dimensional atoms, too. Also, the effect of size-effects caused by confinement in a finite-size graphene is remain in out of the focus of such studies.  The one-dimensional Coulomb potential with application to a class of quasirelativistic systems, so-called Dirac-Weyl materials, described by matrix Hamiltonians is investigated by C. A. Downing and M. E. Portnoi \cite{Downing}.  In this paper, we consider 1D relativistic atom created by a Coulomb impurity in a gapped graphene nanoribbon (GNR) (see,  Fig.1.). We focus on two cases: The case of free, i.e., unconfined atom and when atom is confined in a 1D box 
(atom-in-box system). THe latter can be realized e.g., by a Coulomb impurity in a finite-length graphene nanoribbon.  Within a toy model described in terms of 1D Dirac equation with  Coulomb potential we compute energy eigenvalues and wave functions for both, subcritical and supercritical nucleus charges.  In the next section we briefly recall electronic properties 
of the one-dimensional relativtic atom considered earlier in the Ref.\cite{Downing} by presenting energy levels. Section III presents derivation of the Dirac equation for 1D relativistic atom confined in a box. Section IV treats the problem of critical charge for both systems. In section V we present analysis of local density states for unconfined atom and atom-in-box. Finally, section VI presents some concluding remarks.


\section{Relativistic one-dimensional hydrogen-like atom}

One-dimensional relativistic atom has been considered in a few papers only \cite{Voronina17,Kylstra97}. Lack of high interest to this system can be explained by the fact that it was rather exotic system. Unlike to its non relativistic counterpart. which can be experimentally realized by putting hydrogen atom into uniform electric field, relativistic 1D atom was far from the experimental realization. However, fabrication of graphene and ghraphene nanoribbons, makes possible to create such an atom by doping Coulomb impurities.
Here, following the Ref.\cite{Downing},  we will briefly recall description of the one-dimensional relativistic atom in terms of the Dirac equation.
The motion of the  electron in 1D atomic is governed by the Dirac equation ($\hbar=v_F=1$) \cite{Downing}
$$
H\left(\begin{array}{c} \phi\\\chi  \end{array}\right) =E\left(\begin{array}{c} \phi\\\chi  \end{array}\right)
$$
where the Dirac Hamiltonian is given by

\begin{equation}
    \hat{H}_1=\left(\begin{array}{cc}0 & \hat{p}_x-im_e \\ \hat{p}_x+im_e & 0 \end{array}\right) + V(x), \label{eq02}
\end{equation}

where $_e$ is  the effective mass of the atomic electron and the momentum operator $\hat{p}_x$ acts along the axis of the effectively 1D system. Here $v_F$ is the Fermi velocity, which replaces the light speed in the Dirac equation \cite{Novikov07,Brey,Castro09}.
To avoid the problem of "fall to the center" and take into account the finite-size of the atomic nucleus \cite{Greiner85,Popov01}, the Coulomb potential $V(x)$ can be chosen as \cite{Downing}
\begin{equation}
    V(x) =-\frac{\alpha}{a+|x|}, \lab{shift}    
\end{equation}

or in regularized form (that also takes into account the finite-size  of the atomic nucleus)
\begin{equation} V(x)=\left\{ {\begin{array}{l}
 -\frac{2\alpha}{|d|} \,\,\,\,\,\,\,\,\,\,|x|\leq d,\,\,\, \\
 {-\frac{\alpha}{|x|},\,\,\,\,\,\,\,\,\,\, |x|> d} \\
\end{array}} \right. \lab{reg} \end{equation}

with $\alpha$ being the charge of the nucleus. In what follows, we will focus on shifted Coulomb potential given by Eq.~\re{shift}\\
Using the following unitary transform $H' = U^+ HU,$  where
$$
U=\frac{1}{\sqrt{2}}\left(\begin{array}{cc}
    1 & 1 \\
    1 & -1
\end{array}\right),
$$ 
the Dirac equation can  be rewritten  as
\begin{equation}
    \left(\begin{array}{cc} i \partial_x & -m_e \\ m_e & -i \partial_x \end{array}\right) \left(\begin{array}{c} \phi\\\chi  \end{array}\right)= \left(E_n+\frac{\alpha}{a+|x|}\right)\left(\begin{array}{c} \phi\\\chi  \end{array}\right) \label{eq03}
\end{equation}

The solution of Eq.\ref{eq02}  can be found in two regions \ci{Downing}: In region I ( $x<0$) and in region II ( $x>0$).
For region I we have
\begin{equation}
    \Psi_I=\frac{c_I}{\sqrt{a}}\left(\begin{array}{c}\frac{\kappa_n+iE_n}{m}W_{\mu,\nu+1}(2\kappa_n(a-x))\\W_{\mu,\nu}(2\kappa_n(a-x))\end{array}\right), \lab{sol1}
\end{equation}

and for region II the solution can be written as
\begin{equation}
    \Psi_{II}=\frac{c_{II}}{\sqrt{a}}\left(\begin{array}{c}W_{\mu,\nu}(2\kappa_n(a+x))\\-\frac{\kappa_n+iE_n}{m}W_{\mu,\nu+1}(2\kappa_n(a+x))\end{array}\right).\lab{sol2}
\end{equation}
Here $\kappa_n=\sqrt{m_e^2-E_n^2}$, $\mu=\frac{E_n\alpha}{\kappa_n}$, $\nu=i\alpha-\frac{1}{2}$
and $W_{\mu,\nu}(\xi)$ is the Whitaker function  \ci{Abramowitz}.

 A  secular equation for finding eigenenergies  can be  derived   by matching the two solutions  at $0$ \ci{Downing}:
\begin{equation}
    \frac{m_e}{\kappa_n+iE_n}\frac{W_{\mu,\nu}(2\kappa_n a)}{W_{\mu,\nu+1}(2\kappa_n a)}=\pm i \label{tr_eq}
\end{equation}


\begin{figure}[t!]
\centering
\includegraphics[totalheight=0.035\textheight]{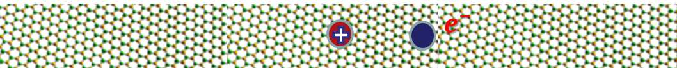}
 \caption{Relativistic 1D atom in narrow graphene nanoribbons} \label{fig0}
\end{figure}




\begin{table}[h!]
\centering {\bf Table 1}. The energy spectrum of  a 1D relativistic hydrogen-like atom at the different values of the nuclear charge $\alpha$ ($m_e\approx0.1$ eV, $d\approx 5.8$ nm). 
\begin{center}
\begin{tabular}{|c|c|c|c|}
\hline \hspace{2mm}n\hspace{2mm} & \hspace{4mm} $\alpha=\frac{300}{137}$
\hspace{4mm} & \hspace{4mm} $\alpha=\frac{600}{137}$ \hspace{4mm} & \hspace{4mm}
$\alpha=\frac{900}{137}$ \hspace{4mm}
\\[3pt]
\hline
1 & -0.030719444 & -0.095148903 & -0.112548882	\\	
\hline
2 & 0.02174156477 & -0.047980966 & -0.076362607	\\	
\hline
3 & 0.052255502 & -0.013454173 & -0.046711014	\\	
\hline
4 & 0.070741651 & 0.012595917 & -0.022722458	\\	
\hline
5 & 0.082230724 & 0.031941611 & -0.002929169	\\	
\hline
6 & 0.089952241 & 0.046904806 & 0.013265771		\\
\hline
7 & 0.095147225 & 0.058291270 & 0.026745874	\\	
\hline
8 & 0.098918098 & 0.067319956 & 0.037914133		\\
\hline
9 & 0.101618468 & 0.074365892 & 0.047302599		\\
\hline
10 & 0.103692415 & 0.080084622 & 0.055178818	\\	
\hline
\end{tabular}
\end{center}
\end{table}

For the truncated Coulomb potential given by Eq~\re{reg} One can write the solution in three regions\cite{Downing} . with the Coulomb 
In the exterior regions ($|x|>d / 2$) I and II,  the solutions are given by Eqs. \re{sol1} and \re{sol2}.
For interior region III($|x| \leq d / 2$), the solutions can be written as \cite{Downing}
\begin{equation}
\Psi_{\mathrm{III}}(x)=\frac{c_{\mathrm{III}}}{\sqrt{d}}\left(\begin{array}{c}
\sin (k x) \\
f_{1}(x)
\end{array}\right)+\frac{c_{\mathrm{IV}}}{\sqrt{d}}\left(\begin{array}{c}
\cos (k x) \\
f_{2}(x)
\end{array}\right)
\end{equation}
where \cite{Downing}
\begin{equation}
\left(\begin{array}{l}
f_{1}(x) \\
f_{2}(x)
\end{array}\right)=\frac{k}{m_e}\left(\begin{array}{c}
\cos (k x) \\
-\sin (k x)
\end{array}\right)+\frac{E+2 \alpha / d}{i m_e}\left(\begin{array}{c}
\sin (k x) \\
\cos (k x)
\end{array}\right)
\end{equation}
and  a "new wave number" is given as         
$k=\sqrt{\left(E+2 \alpha / d\right)^{2}-m_e^{2}}>0$.

\begin{figure}[t!]
\centering
\includegraphics[totalheight=0.2\textheight]{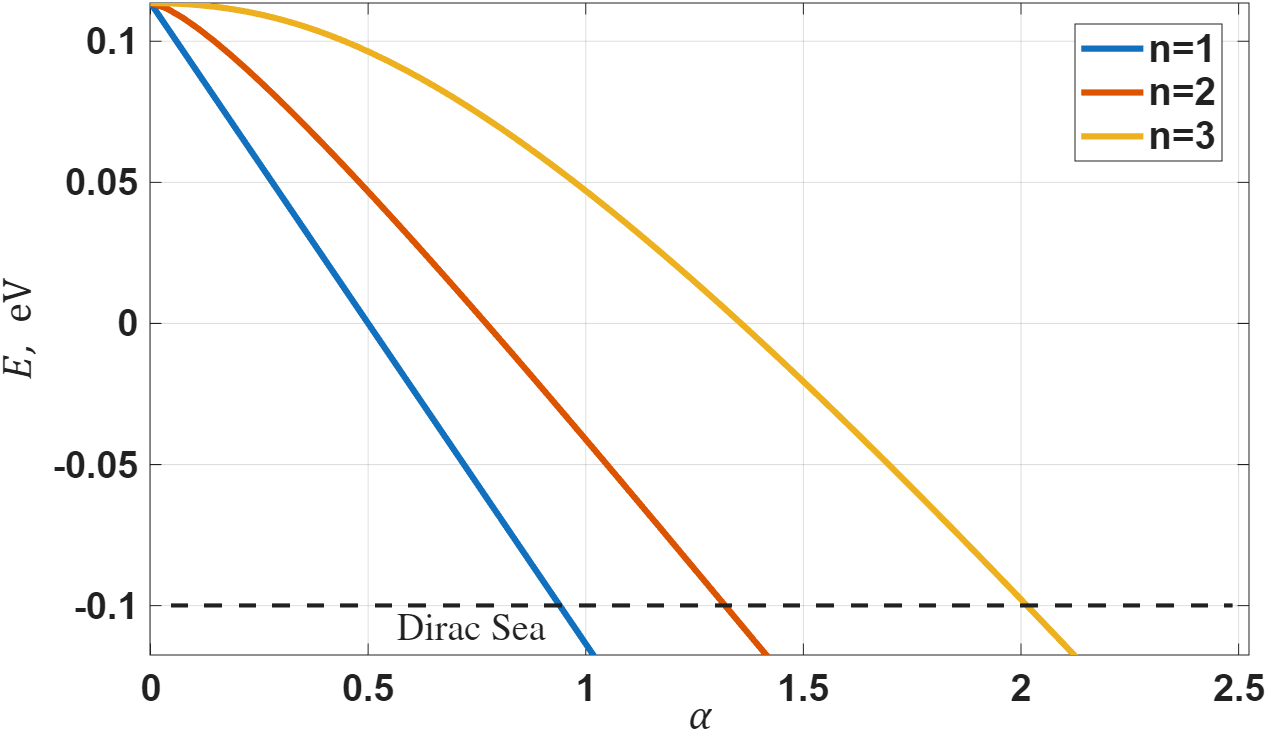}
 \caption{First three lowest energy levels vs nucleus charge, $\alpha$ for unconfined atom for $m_e\approx0.1$ eV} \label{fig1}
\end{figure}

 In Table 1 the first 10 energy levels  of 1D relativistic atom at different values of the atomic nucleus charge, $\alpha$ computed using the  truncated potential given by Eq.~\re{reg} are presented. THe spectrum contains both, negative and positive values energy levels and absolute value of the increases as the number of the level becomes higher. Also, as larger the value of the charge, as higher the number of negative energy levels in the spectrum.

\section{One-dimensional hydrogen atom confined in a box}

An interesting case one can consider in the context of one-dimensional (artificial) relativistic atom is its confined counterpart, i.e. 1D-atom-in-box system. Such system can be formed by Coulomb impurities in a finite-length GNR. Due to the confinement, the spectral and other quantum properties of this system is considerably different than those of the free (unconfined) atom. Such a difference is caused by the boundary conditions to be imposed for the Dirac equation. Studying the role of confinement induced effects in atoms at the quantum level and their macroscopic manifestations is of importance for tuning the physical properties of nanoscale structures and materials. 
Such  system can be described in terms of the Dirac equation with the 1D Coulomb potential for which the box boundary conditions are imposed. Box boundary conditions for one-dimensional Dirac equation have been derived in the Ref.\ci{Alonso}. Here we use the same boundary conditions for our relativistic atom-in-1D-box system.

\begin{figure}[t!]
\centering
\includegraphics[totalheight=0.21\textheight]{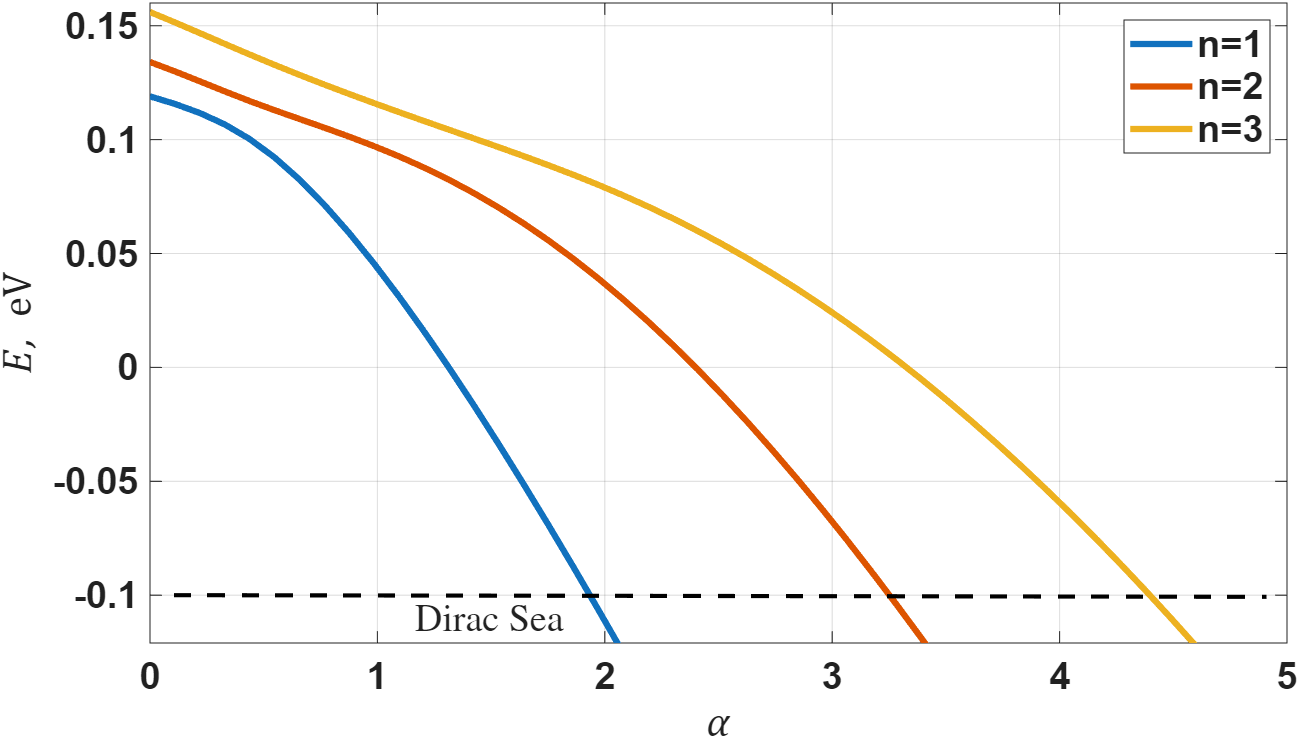}
 \caption{First three energy levels vs $\alpha$ for confined atom at fixed value of parameters, $L\approx58$ nm, $d\approx5.8$ nm,  $m_e\approx0.1$ eV} \label{fig2}
\end{figure}

The Dirac equation for the 1D hydrogen atom confined in a hard wall box can be written as
\begin{equation}
 \left(\begin{array}{cc} i \partial_x +V(x)  & -m_e \\ m_e & -i \partial_x +V(x) \end{array}\right) \left(\begin{array}{c}\phi\\\chi\end{array}\right) = E\left(\begin{array}{c}\phi\\\chi\end{array}\right) \label{DE01} 
\end{equation}

where $V(x)$ is the Coulomb potential of the atomic nucleus given by Eq. \re{reg} or \re{shift}. The components of the 
Dirac spinor $\phi$ and $\chi$ fulfill the box boundary conditions given by 
 \ci{Alonso} 
 $$\phi_1(0)=\phi_1(L)=0.$$

Different versions of the box boundary conditions for the Dirac equation can be found in \ci{Alonso,Alonso02}.  Eq.\re{DE01}, can be solved by expanding $\Psi(x)$ in terms of complete set of the eigenfunctions $\psi_n(x)$ of the Dirac equation for 1D box:
\begin{equation}
    \Psi(x)=\sum C_n\psi_n(x), \label{wf03}
\end{equation}
where $\psi_n(x)$ are given as \ci{Alonso}
$$
\psi_n(x) =A \left(  \begin{array}{c}
     i\sin kx  \\
     \frac{k}{\varepsilon+m_e}\cos kx 
\end{array}  \right)
$$
with $A$ being the normalization coefficient,  and $k_n=\pi n/L$.

Inserting \re{wf03} into Eq. \re{DE01},  and using orthonormality condition, $\int \psi^\dagger(x)_n\psi_m(x)dx=\delta_{nm}$, one can get the following algebraic system:
\begin{equation}
    C_n\varepsilon_n-\sum_mC_mV_{nm}=EC_n
\end{equation}
where $\varepsilon_n =\sqrt{k_n^2+m_e^2}, \;\;$ $V_{nm}=\int\psi^\dagger_n(x)V(x)\psi_m(x)dx.$

Then  the energy levels of relativistic 1D hydrogen atom confined in a hard wall box can be found by diagonalizing the following matrix:
\begin{equation}
    \left(\begin{array}{cccc} \varepsilon_1-V_{11} & -V_{12} & \cdots &-V_{1N} \\ -V_{21} & \varepsilon_2-V_{22} & \cdots & -V_{2N} \\ \vdots & \vdots &\ddots &\vdots \\ -V_{N1} &-V_{N2} &\cdots & \varepsilon_N-V_{NN} \end{array}\right). \label{mat}
\end{equation}

\newpage

\begin{widetext}
    
\begin{table}[h!]
\centering {\bf Table 2.} First ten energy levels (eV) of the 1D relativistic hydrogen-like atom in a box with the nuclear charge $\alpha=300/137$ at different values of the box size, $L$ ($m_e\approx0.1$ eV, $d\approx5.8$ nm). 
\begin{center}
\begin{tabular}{|c|c|c|c|c|}
\hline \hspace{5mm}n\hspace{5mm} & \hspace{5mm} $L\approx58$ nm
\hspace{5mm} & \hspace{5mm} $L\approx290$ nm \hspace{5mm} & \hspace{5mm}
$L\approx580$ nm \hspace{5mm}& \hspace{5mm}
Free atom \hspace{5mm}
\\[3pt]
\hline 1 & -0.144606701  & -0.144607596	&	-0.144610462  & -0.030719444 \\	
\hline
2 & 0.0201893699	&	0.0201890182	&	0.0201883026  &  0.021741564	\\
\hline
3 & 0.0706447984	&	0.0705055946	&	0.0705053697	&  0.052255502 \\
\hline
4 & 0.0943246017	&	0.0898780206	&	0.0898779261	& 0.070741651 \\
\hline
5 & 0.118935682	&	0.0988856049	&	0.098885558  & 0.082230724	\\
\hline
6 & 0.147358445	&	0.10367535	&	0.10367532 &  0.089952241	\\
\hline
7 & 0.177992899	&	0.106491935	&	0.106486201  &  0.095147225	\\
\hline
8 & 0.209965640	&	0.108455203	&	0.108262726	&  0.098918097 \\
\hline
9 & 0.242806269	&	0.110565384	&	0.109451571 & 0.101618467	\\
\hline
10 & 0.276240924	&	0.113186888	&	0.110285874 &  0.103692415	\\

\hline
\end{tabular}
\end{center}
\end{table}

\end{widetext}

\begin{figure}[t!]
\centering
\includegraphics[totalheight=0.23\textheight]{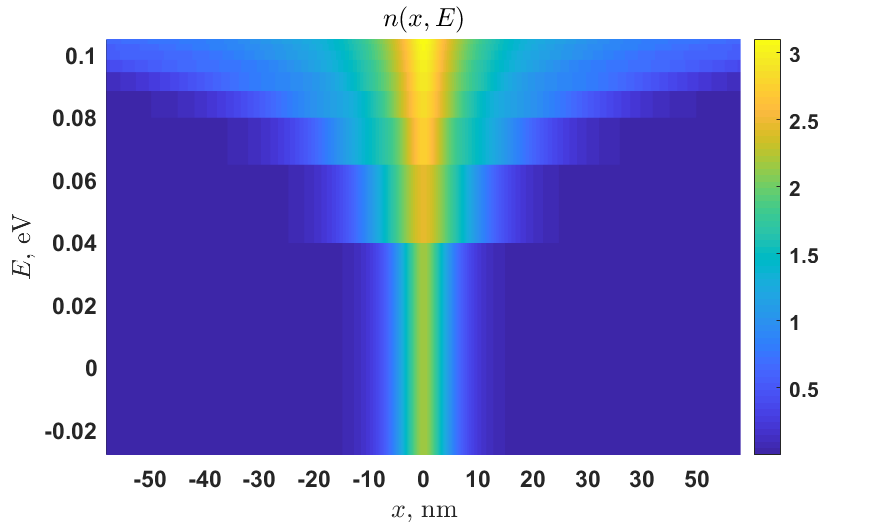}
 \caption{Coordinate and energy dependence of the local density of states of the unconfined atom at fixed value of parameters, $m_e\approx0.1$ eV, $a\approx5.8$ nm, $\alpha=300/137$} \label{fig3}
\end{figure}

First 10 energy levels of the atom for $\alpha=300/137$ at different values of the box size are presented in Table 2 and compared with those of unconfined atom. 
Feasible dependence of the energy level on the box size starts from the fourth level. Big difference between the energy levels for confined and free atoms implies crucial role of the confined on the energy spectra. As smaller the size of the box, as higher the binding energy. Table 3 presents energy levels of the atom-in-box system at different values of the charge for the fixed size of the box.

\section{Critical charge}
An important characteristics of relativistic charged particle interacting with strong external fields is so-called critical field strength.
For strong enough fields, i.e. by increasing field strength up to certain value, the binding energy level of the electron can descend the border of the Dirac sea. For Coulomb field, the critical field strength is determined by the critical charge, .e., the value of the nucleus charge, at which a given energy level reaches the boundary of the Dirac sea.
Calculation of the critical charge is of importance for estimating threshold value of the fields causing the so-called supercritical phenomena, such as 
electron-positron pair creation, vacuum polarization and Klein tunneling.

To compute the critical charge for our 1D atom, we use regularized Coulomb potential given by Eq.\re{reg}. Requiring  the continuity of the wave-function components at $x=\pm d / 2$ one can obtain the following transcendental equation  \ci{Downing}:
\begin{equation}
1-\lambda_{+} / \lambda_{-}=0, \lab{transeq1}
\end{equation}
where
$$
\lambda_{\pm}=\frac{\frac{k}{m_e \tau_{\pm}} \tan \left(\frac{k d}{2}\right) \pm \eta_{\pm}}{\eta_{\pm} \tan \left(\frac{k d}{2}\right) \mp \frac{k}{m_e \tau_{\pm}}} \\
$$
$$
\eta_{\pm}=i\left(\frac{E+2 \alpha / d}{m_e \tau_{\pm}}\right) \mp 1 \\
$$
and

$$
\tau_{\pm}=\left(\frac{\kappa+i E}{m_e} \frac{W_{\mu, \nu+1}(\kappa d)}{W_{\mu, \nu}(\kappa d)}\right)^{\pm 1}
$$

\begin{table}[h!]
\centering {\bf Table 3.} The energy levels (eV)  of  1D relativistic hydrogen-like atom in box at different values of nucleus charge, $\alpha$ for fixed box size, $L\approx580$ nm ($m_e\approx0.1$ eV, $d\approx5.8$ nm). 
\begin{center}
\begin{tabular}{|c|c|c|c|}
\hline \hspace{2mm}n\hspace{2mm} & \hspace{8mm} $\alpha=\frac{300}{137}$
\hspace{4mm} & \hspace{4mm} $\alpha=\frac{600}{137}$ \hspace{4mm} & \hspace{4mm}
$\alpha=\frac{900}{137}$ \hspace{4mm}
\\[3pt]

\hline
1 & -0.144610463	&	-0.566672502	&	-1.020465285	\\
\hline
2 & 0.0201883026	&	-0.266080778	&	-0.650270091	\\
\hline
3 & 0.0705053697	&	-0.0981445371	&	-0.384133945	\\
\hline
4 & 0.0898779261	&	-0.0141381104	&	-0.217885182	\\	
\hline
5 & 0.098885558	&	0.0317478782	&	-0.114316716	\\
\hline
6 & 0.10367532	&	0.0582210615	&	-0.047316196	\\	
\hline
7 & 0.106486201	&	0.0743348136	&	-0.0031731094	\\	
\hline
8 & 0.108262725	&	0.0846363461	&	0.0266370592	\\	
\hline
9 & 0.109451571	&	0.0915156952	&	0.0472487250	\\	
\hline
10 & 0.110285874	&	0.0962878473	&	0.0618390796	\\
\hline
\end{tabular}
\end{center}
\end{table}

\begin{figure}[t!]
\centering
\includegraphics[totalheight=0.23\textheight]{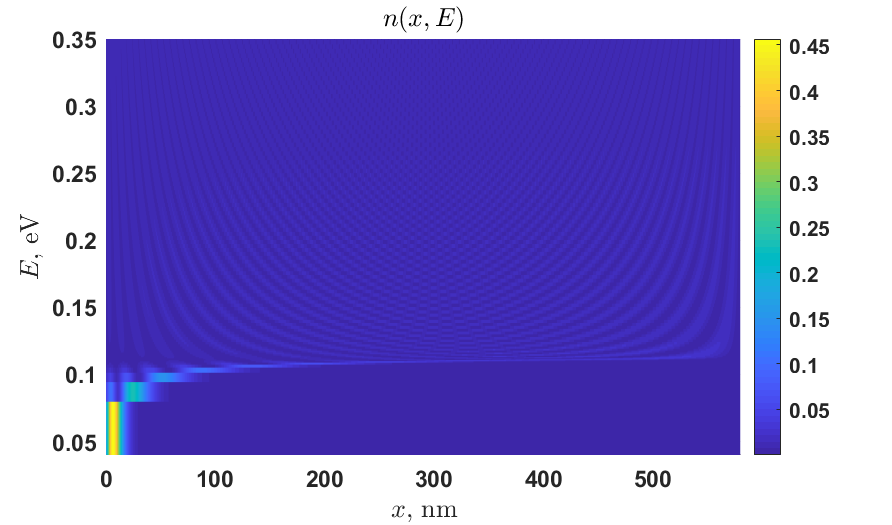}
 \caption{Coordinate and energy dependence of the local density of states of the confined atom at fixed value of parameters, $L\approx580$ nm, $m_e\approx0.1$ eV, $a\approx5.8$ nm, $\alpha=300/137$} \label{fig4}
\end{figure}

In the system of units, $ \hbar=m_e=v_F=1$, Eq. \re{transeq1}, can be written as
 \begin{equation}
 \begin{split}
k(\tau_{+}+\tau_{-})\Bigl(\tan^2(k/2)-1 \Bigl)+2\tan(k/2) \\
\times \biggl (i(E+2\alpha)(\tau_{-}-\tau_{+})-2)\biggl)=0
 \end{split}
 \lab{transeq2}
 \end{equation}
where $ k=\sqrt{(E +2\alpha )^2-1}$, $\tau_{+}=(\kappa+i E)\frac{W_{\mu,\nu + 1}(\kappa)}{W_{\mu,\nu}(\kappa)}$, 
$\tau_{-}=\frac{1}{\tau_{+}}$, $ \mu=\frac{E \alpha }{\sqrt{1- E^2}} $, $ \nu=i\alpha-\frac{1}{2} $ and $ \kappa=\sqrt{1-E^2} $.

The critical charge can be found from Eq.\re{transeq2} by putting  $E=-1$. 
The values of the  critical charge, $\alpha_{cr}$ for ten lowest energy levels are presented in the Table 4.\\
The dependence of the critical charge on the size of the box is presented in Table 5. For confined atom, the values of $\alpha_{cr}$ are much larger than those for the free atom. As becomes smaller the size of the box, as larger the value of $\alpha_{cr}$.  

Fig. 3 presents first three energy levels of the "atom + 1D box" system as a function of the charge. Critical charge corresponds to the crossing point of the curves with -1.

Fig. 2 presents first three (lowest) energy levels vs nucleus charge, $\alpha$. The value of $\alpha$, where $E(\alpha)$ crosses $-1$ corresponds to the critical charge.

\begin{figure}[t!]
\centering
\includegraphics[totalheight=0.23\textheight]{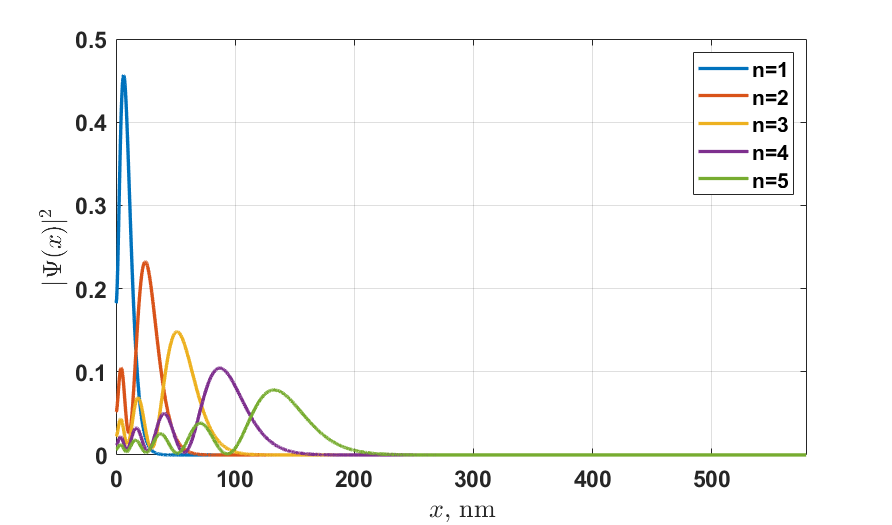}
 \caption{Coordinate dependence of the probability of confined atom  for initial few levels at fixed value of parameters, $L\approx580$ nm, $m_e\approx0.1$ eV, $a\approx5.8$ nm, $\alpha=300/137$} \label{fig6}
\end{figure}

\section{Local density of states (LDOS) for unconfined 1D relativistic Coulomb problem for Dirac equation}

Due to the fact that graphene is an atomically thin membrane, it can be easily studied with
Scanning Tunneling Microscopy (STM) measurements. Especially, this concerns Coulomb impurity effects, which   can be studied with atomic resolution and the local spectrum can be obtained by STM spectroscopy \ci{Uchoa09,Carvalho14,Peres09}. Also, atomic manipulation can also be performed using STM. The main characteristic that can be measured using STM
is the local density of states, which is defined by \ci{Uchoa09,Carvalho14,Peres09}

\begin{equation}
    n(x,E)=\sum_j|\Psi_j(x)|^2\delta(E-\varepsilon_j)
\end{equation}

In Fig.\re{fig3} the local density of states (LDOS) of unconfined atom is plotted as a function of coordinate and energy at fixed value of parameters, $m_e\approx0.1$ eV, $a\approx5.8$ nm, $\alpha=300/137$. Localization around $x=0$ can be clearly seen from the plot. Such a localization and symmetry with respect to the origin of coordinate is caused by the choice of 1D Coulomb potential and the shape of the probability density (see, e.g., \ci{Downing}), which is also symmetric  with respect to $x=0.$ \\
Fig.\re{fig4} presents similar plot for confined atom for the values of parameters,  $L\approx580$ nm, $m_e\approx0.1$ eV, $a\approx5.8$ nm, $\alpha=300/137$. 
Localization around $x=0$ can be also observed in this case, although there is no symmetry with respect to the origin of coordinate. The latter is caused by the shape of probability density, $|\psi(x)|^2$ which is localized in the right hand of $x=0$ (see, Fig.6). The value of $n(x,E)$ is much higher for unconfined atom that makes also localization stronger for this case. 

\section{Conclusions}

We considered a toy model of a one-dimensional relativistic hydrogen-like atom that can be realized through Coulomb impurities in Dirac materials, such as graphene nanoribbons. The study focuses on the spectral properties of the system and on the critical charge responsible for the "diving" of a given energy level into the Dirac sea. A truncated Coulomb potential is employed to account for the finite size of the atomic nucleus.
Both the free-atom case and the "atom-in-a-hard-wall-box" configuration are investigated. For each case, the energy 

\newpage
\begin{widetext}

\begin{table}[h!]
\begin{center}
\centering {\bf Table 4.} The critical charge by energy states of a $1 \mathrm{D}$ relativistic hydrogen-like unconfined atom ($m_e\approx0.1$ eV, $d\approx 5.8$ nm). 
\begin{tabular}{|c|c|c|c|c|c|c|c|c|c|c|}
\hline
\hspace{5mm} $\mathrm{n}$ \hspace{5mm} &  \hspace{5mm} 1 \hspace{5mm} & \hspace{5mm} 2 \hspace{5mm} & \hspace{5mm} 3 \hspace{5mm} & \hspace{5mm} 4 \hspace{5mm} & \hspace{5mm} 5 \hspace{5mm} & \hspace{5mm} 6 \hspace{5mm} & \hspace{5mm} 7 \hspace{5mm} & \hspace{5mm} 8 \hspace{5mm} & \hspace{5mm} 9 \hspace{5mm} & \hspace{5mm} 10 \hspace{5mm} \\
\hline
$\alpha_{cr}$ & 0.997 & 1.3927 & 2.0969 & 2.801 & 3.428 & 4 & 4.553 & 5.087 & 5.596 & 6.095 \\
\hline
\end{tabular}
\end{center}
\end{table}

\begin{table}[h!]
\begin{center}
\centering {\bf Table 5.} The critical charge (for  $1 \mathrm{D}$ relativistic atom confined in a box) for different energy levels  at different values of the box size ($m_e \approx 0.1$ eV, $d \approx 5.8$ nm).
\begin{tabular}{|c|c|c|c|c|c|c|c|c|c|c|}
\hline
 \diagbox[width=6em,height=2em]{\hfill $L$ \hfill}{\hfill $\mathrm{n}$ \hfill}  &  \hspace{5mm} 1 \hspace{5mm} & \hspace{5mm} 2 \hspace{5mm} & \hspace{5mm} 3 \hspace{5mm} & \hspace{5mm} 4 \hspace{5mm} & \hspace{5mm} 5 \hspace{5mm} & \hspace{5mm} 6 \hspace{5mm} & \hspace{5mm} 7 \hspace{5mm} & \hspace{5mm} 8 \hspace{5mm} & \hspace{5mm} 9 \hspace{5mm} & \hspace{5mm} 10 \hspace{5mm} \\
\hline
$5.8$ nm & 2.432 & 4.31 & 6.173 & 8.03 & 9.886 & 11.737 & 13.591 & 15.446 & 17.3 & 19.155 \\
\hline
$11.6$ nm & 2.062 & 3.505 & 4.855 & 6.171 & 7.478 & 8.787 & 10.1 & 11.41 & 12.723 & 14.036 \\
\hline
$29$ nm & 2.0131 & 3.3535 & 4.5223 & 5.5756 & 6.572 & 7.532 & 8.471 & 9.401 & 10.329 & 11.256 \\
\hline
$58$ nm & 2.013 & 3.353 & 4.521 & 5.571 & 6.56 & 7.504 & 8.413 & 9.295 & 10.154 & 10.994 \\
\hline
\end{tabular}
\end{center}
\end{table}

\end{widetext}

spectra and critical charge are calculated. The role of spatial confinement in the electronic energy spectrum is analyzed through a comparison of the free and confined systems. In the confined case, the energy levels are significantly smaller than those of the free atom. Reducing the confinement size (i.e., decreasing the box size) leads to an increase in the energy levels.
Confinement also increases the value of the critical charge: the smaller the box size, the larger the critical charge. The proposed one-dimensional relativistic atomic model provides an effective framework for the experimental realization of Coulomb impurities in graphene and for STM-based measurements of their spectral properties and critical charge. A more realistic description should be based on the shrinking-limit two-dimensional Dirac equation with zigzag or armchair boundary conditions. Such an investigation is left for future research.

\acknowledgements
We acknowledge funding by the Grant REP-05032022/235 ("Ultrafast phenomena and vacuum effects in relativistic artificial atoms created in graphene"), funded under the MUNIS Project, supported by the World Bank and the Government of the Republic of Uzbekistan Also, the work of DM was supported in part by the grant of the Innovations Agency under the Ministry of Higher Education, Research and Innovations (FL-8824063336).

\end{document}